# Controlled Interlayer Exciton Ionization in an Electrostatic Trap in Atomically Thin Heterostructures

Andrew Y. Joe[1,†,#], Andrés M. Mier Valdivia[2,†], Luis A. Jauregui[3], Kateryna Pistunova[1], Dapeng Ding[1,4], You Zhou[1,4,5], Giovanni Scuri[1], Kristiaan De Greve[1,4], Andrey Sushko[1], Bumho Kim[6], Takashi Taniguchi[7], Kenji Watanabe[8], James C. Hone[6], Mikhail D. Lukin[1], Hongkun Park[1,4] and Philip Kim[1,2*]

[1]Department of Physics, Harvard University, Cambridge, 02138, MA, USA.
[2]John A. Paulson School of Engineering and Applied Sciences, Harvard University, Cambridge, 02138, MA, USA.
[3]Department of Physics, University of California, Irvine, 9269, CA, USA.
[4]Department of Chemistry and Chemical Biology, Harvard University, Cambridge, 02138, MA, USA.
[5]Department of Materials Science and Engineering, University of Maryland, College Park, 20742, MD, USA.
[6]Department of Mechanical Engineering, Columbia University, New York, 10027, NY, USA.
[7]International Center for Materials Nanoarchitectonics, National Institute for Materials Science, 1-1 Namiki, 305-0044, Tsukuba, Japan.
[8]Research Center for Electronic and Optical Materials, National Institute for Materials Science, 1-1 Namiki, 305-0044, Tsukuba, Japan.

[†] These authors contributed equally to this work.
[#] Currently at Department of Physics and Astronomy, University of California, Riverside, CA
[*] Corresponding author: pkim@physics.harvard.edu

**Abstract:** Atomically thin semiconductor heterostructures provide a two-dimensional (2D) device platform for creating high densities of cold, controllable excitons. Interlayer excitons (IEs), bound electrons and holes localized to separate 2D quantum well layers, have permanent out-of-plane dipole moments and long lifetimes, allowing their spatial distribution to be tuned on demand. Here, we employ electrostatic gates to trap IEs and control their density. By electrically modulating the IE Stark shift, electron-hole pair concentrations above $2\times10^{12}$ cm$^{-2}$ can be achieved. At this high IE density, we observe an exponentially increasing linewidth broadening indicative of an IE ionization transition, independent of the trap depth. This runaway threshold remains constant at low temperatures, but increases above 20 K, consistent with the quantum dissociation of a degenerate IE gas. Our demonstration of the IE ionization in a tunable electrostatic trap represents an important step towards the realization of dipolar exciton condensates in solid-state optoelectronic devices.

# Introduction

Semiconductor van der Waals (vdW) heterostructures offer a unique platform where strong light-matter interactions enable novel device capabilities that expand our ability to study fundamental mesoscopic phenomena. As their building blocks, monolayer transition metal dichalcogenides (TMDs) provide a synergy of electronic and photonic degrees of freedom that manifest as excitons, optically generated bound electron-hole pairs, with large binding energies[1] and spin-valley locking properties[2,3]. In type-II TMD heterostructures, such as $MoSe_2/WSe_2$, electrons and holes are localized on separate 2D quantum well layers. This band alignment leads to the formation of IEs with permanent out-of-plane dipole moments[4–7], extended lifetimes[4,8,9], and long interaction-driven diffusion lengths[4,7,10], while maintaining strong binding energies[11–13]. These properties also make IEs ideal for the design of excitonic devices with engineered spatial emission profiles and controllable IE density with electrostatic gates.

The generation and control of large IE concentrations is desirable to explore the rich physics predicted by the excitonic phase diagram. In GaAs double quantum wells, indirect excitons have been shown to condense into a degenerate state[14–16] that is expected to be a superfluid, providing a route to dissipationless optoelectronic devices. While increasing exciton density is generally favorable for condensation, increasing wavefunction overlap eventually leads to complete screening of the IE binding energy at the Mott density[17–19]. Even below this critical condition, a fraction of the IEs can be thermally ionized at a finite temperature, enhancing the screening effect. The previous theoretical studies showed that this density dependent screening leads to a runway IE dissociation process near the ionization threshold[20–25]. In TMD heterostructures, the orders of magnitude larger binding energies allow the system to sustain larger IE populations before ionization effects become relevant[9]. This property opens the door to the realization of electrically tunable high-temperature condensates[13,26,27].

Electrostatics traps are a useful tool for exploring the IE phase diagram. They allow us to deterministically localize IEs and increase their density, bypassing the need for high laser powers that lead to heating or nonlinear Auger recombination[28,29], and potentially even damage to the sample. Furthermore, the control over their diffusion dynamics opens the avenue for reliably designing excitonic devices, as other trapping approaches like moiré localization depend sensitively on sample homogeneity[30–33]. Separate control of the IE density and temperature paves the way towards constructing tunable coherent optoelectronic devices and studying correlated phenomena emerging from them.

# Results

**Electrostatic tunability of interlayer exciton emission**

In this article, we design a multi-gated structure to create electrostatic traps that allow us to generate high densities of IEs, control their spatial distribution, and determine their critical dissociation density as a function of temperature. All the data shown is from device D1 unless otherwise stated. Figure 1(a) shows a schematic of the device design. A high quality, h-BN encapsulated $MoSe_2/WSe_2$ heterostructure is fabricated with matching top and bottom metal gates (Methods). A matching set of grounded ~250 nm-wide stripe gates maintains a constant IE energy in the trap region. On either side of the stripe are two sets of top and bottom gates separated by ~500 nm that control the IE energy in the outside region. Applying voltages to the outer gates creates a neighboring electric field, $E_{og} = \frac{\varepsilon_{h-BN}}{\varepsilon_{TMD}} \frac{V_{tg} - V_{bg}}{t_{total}}$, where $\varepsilon_{h-BN}$ = 3.9, $\varepsilon_{TMD}$ = 7.2 are the h-BN and TMD dielectric constants, $V_{tg}$ and $V_{bg}$ are the top and bottom gate voltages, and $t_{total}$ is the total h-BN thickness. Applying positive $E_{og}$ raises the IE energy in the outer region due to the linear Stark effect[4]. We note that the intralayer exciton energies are not affected in this gating configuration since their dipole moments are in the TMD plane[4]. The electric field for the center stripe gates is approximately zero since these are grounded. The relative field difference between the stripe gate region and the outer area creates a confinement potential along the former so that a high IE density can accumulate in the trap without diffusing away (red curve in Fig. 1(b)). Fig. 1(c) shows an optical image of the device detailing the spatial region for the experiment.

We demonstrate our gate modulated trap by performing scanning photoluminescence (PL) spectroscopy at various $E_{og}$ across the stripe gates (Methods). All PL measurements are performed at $T = 4$ K, unless otherwise stated, with an excitation wavelength of 660 nm (1.87 eV), and a diffraction limited spot. In Figure 1(d), we measure the PL spectra as a function of $E_{og}$ at a power ($P$) of 50 µW at the edge of the stripe gates. As $E_{og}$ increases, a second peak splits off, with only the higher energy peak shifting linearly with the external field. We extract a dipole separation of $d \approx 0.6$ nm from the slope of the energy shift $\Delta E = -edE_{og}$ ($e$ is the elementary charge), corresponding to the expected interlayer separation between the electron and hole wavefunctions localized on the TMD layers. In Figures 1(e)-(g), we excite the heterostructure with $P = 500$ µW at the center of the stripe gate and measure the PL spectra at a transverse distance $x$ away from the center of the stripe gate for three representative fields, $E_{og} = 0$ V/nm and ±0.114 V/nm. When $E_{og} = 0$ V/nm, the PL energy is constant at ~1.38 eV across the spatial linecut, consistent with the expected electrostatic profile. When $E_{og} = 0.114$ V/nm, the PL energy in the outer gate region rises to ~1.44 eV, consistent with the gate dependent spectra, while the stripe gate region emits lower energy. Thus, we conclude that the higher energy peak originates from IEs outside the trap, and we identify the lower energy peak as the trapped IEs. Regardless of the trapping potential, we observe that the higher-energy free IE emission extends nearly 3 µm away from the center of the stripes. The observed long diffusion lengths of IEs in TMDs is caused by dipolar repulsion[4]. However, we note that the high energy emission is darker when the confinement potential is active, hindering IE diffusion perpendicular to the stripe, while favoring localization in the trap. We also demonstrate an anti-trap potential when $E_{og} = -0.114$ V/nm (Fig. 1g), where the overall emission is darker because the electrostatic profile helps IEs diffuse away from the excitation spot (see also Supplementary Fig. S1).

**Spatial control over interlayer exciton dynamics**

To understand the diffusive behavior under confinement, we take spectrally filtered PL intensity maps at the IE energy. Figures 2(a)-(c) show the spatially dependent IE PL emission for anti-trap, flat, and trap potentials when the excitation spot is fixed near the edge of the stripe gate for $P = 550$ µW (red circle in Fig. 2(a)). For $E_{og} = -0.023$ and 0 V/nm, we observe emission around the laser spot and at uncontrolled disorder-related bright spots. In the absence of the externally applied trapping potential, IEs diffuse and localize in these naturally occurring traps (Supplementary Fig. S2(b)-(c)). However, when $E_{og} = 0.114$ V/nm, we observe uniform and elongated emission along the stripe, implying that the IE diffusion is constrained along it. In Fig. 2(d), we average the emission along the stripe ($y$-direction) and plot the normalized intensity as a function of $x$ perpendicular to the stripe for various $E_{og}$. We observe a narrowing of the spatial IE cloud width when the trap is active and a broadening when an anti-trap is created. In Fig. 2(e), we show three characteristic spatial PL profiles obtained along the linecuts shown in Fig. 2(d), at constant $E_{og}$ corresponding to anti-trap (yellow), no-trap (black), and trap (red) configurations. These profiles exhibit a spatially narrowed (broadened) distribution for $E_{og} = 0.114$ V/nm ($-0.023$ V/nm) compared to 0 V/nm. We perform a similar analysis along the stripe gate and find that the emission is uniform above $E_{og} \approx 0.09$

V/nm (Supplementary Fig. S3), corresponding to a trap depth of ~55 meV. This estimated natural trap energy scale is consistent with reports of defect or strain-trapped IEs in these heterostructures[34,35].

The emission profile across the stripe gate can be described by the balance between the diffusion of laser-generated IEs near the edge of the stripe, and the electrostatic trap (anti-trap) that hinders (favors) exciton transport. We obtain the spatial full-width half-maximum (FWHM) centered at the trap using a Gaussian fit. Fig. 2(f) shows the FWHM as a function of $E_{og}$. We observe that the trap diffusion width saturates near 0.95 μm at $E_{og} \approx 0.04$ V/nm. The FWHM is wider than the expected trap width (~500 nm), which could be explained by imperfect alignment of the top and bottom gates or by convolution with the point spread function of the system (Supplementary Information S.4). We note that the FWHM increases slightly for $E_{og} > 0.12$ V/nm, probably due to a shorter IE lifetime and greater emission outside the trap for larger electric fields[4]. The relative peak PL (normalized to the emission at $E_{og} = 0$ V/nm) also increases rapidly until $E_{og}$ reaches ~ 0.04 V/nm, before increasing at a slower rate. We observe the longest IE lifetime at the center stripe region when a trapping potential is applied, which increases by up to 20% at the largest $E_{og}$ (Supplementary Fig. S5). These signatures indicate stronger confinement at higher IE densities.

**Generation of a high-density trapped electron-hole ensemble**

The PL peak energy shifts as a function of electron-hole pair density due to interactions. In the excitonic limit, strong dipolar repulsion leads to a blueshift for increasing IE density[4,5,9,18,19,36–38]. Similarly, an increasing density of unbound electrons and holes on opposite layers generates an electric field leading to a blueshift[18,39,40]. Fig. 3(a) shows the PL spectra as a function of the excitation power $P$ at the center of the trap (see Fig. 2). Even without trap activation (i.e., $E_{og} = 0$), the PL peak exhibits a strong blueshift as $P$ increases (Fig. 3(a)). This increase of the IE emission energy $\Delta E$ can be described by a balance between the diffusion, generation, and emission of IEs[4]. With increasing $P$ in the excitonic regime, the electron-hole pair density $n_{eh}$ under the stripe gates increases as excitons diffuse from the excitation site, leading to an increase in $\Delta E$. The pair density can be further increased by deepening the confinement potential: a larger $E_{og}$ hinders diffusion out of the trap as in Fig. 2(f), leading to a larger blueshift (Fig. 3(b)) for the same power range compared to the flat potential case.

For a more quantitative estimate of $n_{eh}$, we employ a formula obtained by considering dipolar excitonic interactions using a mean-field approach[39]: $\Delta E = \frac{(n_{eh}-n_0)e^2 d}{\varepsilon_{TMD}\varepsilon_0}$, where $n_0$ is the initial density from which we reference $\Delta E$, and $\varepsilon_0$ is the vacuum permittivity. Although this linear relationship tends to underestimate the pair density[40], we treat $\Delta E$ as an indication of increasing electron-hole concentration in the trap. In Fig. 3(c), we plot the measured $\Delta E$ (left axis) and the estimated $n_{eh}$ (right axis) at the center of the stripe gate region as a function of $P$ with (orange symbols) and without (black symbols) the trap potential activated. We estimate a small initial density $n_0$ for the lowest measured excitation power ($P = 1$ μW) (Supplementary Information S6). Since $n_0$ and the blueshift at low powers are small, this estimation contributes only a small error at higher densities. At low powers ($P < 10^{-4}$ W), we find that $n_{eh}$ behaves similarly with and without a trap potential, increasing slowly with increasing $P$. This behavior suggests that IE diffusion is confined to the trap region, independent of the trap depth, and therefore $n_{eh}$ is limited by the IE generation and recombination rates. However, at high powers ($P > 10^{-4}$ W), the maximum $n_{eh}$ with the trap on is nearly twice as large as with the trap potential off, indicating that the electrostatic profile confines electron-hole pairs that would otherwise diffuse away at the same generation rate. Without a trap, we obtain a maximum $n_{eh} = 1.38 \times 10^{12}$ cm$^{-2}$, in agreement with previous measurements[4]. With the trapping potential, we reach almost twice this density: $n_{eh} = 2.40 \times 10^{12}$ cm$^{-2}$. We note that the enhanced blueshift occurs only for trapped IEs and not for IEs outside of the trap even at $E_{og} = 0.114$ V/nm (Supplementary Fig. S6). This confirms that the blueshift enhancement is a consequence of increasing interactions from higher local electron-hole densities.

In addition to the blueshifts, we observe a broadening of the linewidth, shown by the contour lines of half maximum normalized intensity (Figs. 3(a)-(b)). Figure 3(d) shows the fitted linewidth as a function of power for $E_{og} = 0$ and $0.114$ V/nm. At lower excitation powers, we have a linewidth of ~10 meV in both cases. At higher $E_{og}$, we observe an increase in the linewidth with lower power, similar to the deviations observed in $n_{eh}$ as discussed above. We rule out heating as the main mechanism causing this trend, since a temperature increase due to laser heating would lead to linewidth broadening and peak redshifting regardless of the trap potential set by $E_{og}$ (see S.11).

**Interlayer exciton ionization phase diagram**

The rapid increase of the spectral linewidth at high power invites a more detailed investigation. Figure 4(a) shows the density dependent linewidth $w(n_{eh})$ at different $E_{og}$ and $P$. As discussed in Fig. 3, $n_{eh}$ can vary over a wide range ($10^9$ to $2 \times 10^{12}$ cm$^{-2}$). Interestingly, we find that $w(n_{eh})$ follows a universal curve determined solely by $n_{eh}$, independent of $E_{og}$ and $P$. We find that the linewidth increases exponentially with $n_{eh}$ (inset of Fig. 4(a)), suggesting that we can fit the behavior with the functional form $w(n_{eh}) = w_0 \exp\left(\frac{n_{eh}}{n_{eh}^*}\right)$, where $w_0$ is an intrinsic linewidth at low pair density and $n_{eh}^*$ is a characteristic density for the exponential upturn. We consider a rapid increase in the spectral linewidth for $n_{eh} > n_{eh}^*$ as a signature of the ionization threshold of IEs, equivalent to the Mott transition[9,17,20–24,37,41]. At base temperature, we estimate $n_{eh}^*(4\ K) = 1.30\ (\pm 0.09) \times 10^{12}$ cm$^{-2}$ (Figure 4 (a)). We observed similar behavior exponential behavior in two additional devices, D2 and D3, with different stacking orientations and interlayer h-BN spacing (see Fig. S11 and Supplementary Section S.10 for further discussion). Interestingly, all heterostructures exhibit an $n_{eh}^* \sim 1 - 2.5 \times 10^{12}$ cm$^{-2}$. For D1, we repeated similar measurements at different temperatures $T$ ranging from 4 K to 70 K, where $n_{eh}^*(T)$ can be similarly obtained from the exponentially increasing linewidth (see Fig. 4 (b,c) for representative data at higher temperatures and more data is available in Supplementary Fig. S8). Figure 4(d) shows the normalized linewidth as a function of $n_{eh}$ and $T$. In this plot we also show $n_{eh}^*(T)$ obtained from the exponential fit. For $T \leq 20$ K, we find that $n_{eh}^*(T)$ is constant within our fitting uncertainty, while above 20 K, $n_{eh}^*(T)$ increases with increasing temperature.

The broadening of the spectral linewidth as a function of $n_{eh}$ has been previously observed for indirect excitons in GaAs double quantum wells[17,18,38,41] and for IEs in MoSe$_2$/WSe$_2$ heterostructures[9,37], and is attributed to the dissociation of the IEs. Similarly, we attribute $n_{eh}^*$ to the density driven ionization threshold for IEs. As $n_{eh}$ increases, Coulomb screening becomes more pronounced, reducing the binding energy and gradually increasing the fraction of ionized carriers[21]. At $n_{eh}^*$, the mixed IE-plasma system reaches a threshold and ionization becomes a runaway process[20–25], which can explain the rapid increase of the PL linewidth, although a complete theoretical model of the exponential dependence is beyond the scope of our work. Following this scenario, the temperature dependent behavior of $n_{eh}^*(T)$ shown in Fig. 4(d) is understood by considering the competing effects of screening and bosonic degeneracy. Here, the three different length scales that characterize this competition are the excitonic Bohr radius $a_B$, the mean interpair distance $D = n_{eh}^{-1/2}$, and the excitonic thermal de Broglie wavelength $\lambda_T = \left(\frac{2\pi\hbar^2}{m_{IE} k_B T}\right)^{-\frac{1}{2}}$, where $m_{IE}$ is the total IE mass. In the low temperature and high density regime where $D < \lambda_T$, the quantum statistical nature of the bosonic IE becomes appreciable. The condition $\lambda_T = D$ defines the degeneracy temperature per flavor $T_D = \frac{2\pi\hbar^2}{k_B m_{IE} n_{eh}}$ (red dashed line in Fig. 4(d)), which marks the crossover between the classical (non-degenerate) and quantum (degenerate) IEs. Experimentally, we find that the temperature-dependent ionization threshold $n_{eh}^*(T)$ occurs well below this line for all measured temperatures in our experiment, indicating that the transition we observe is related to the quantum dissociation of the IEs. Below 20 K, we find that $n_{eh}^*(T)$ remains at a nearly constant value of ~1.3×10$^{12}$ cm$^{-2}$. This $n_{eh}^*$ behavior can be understood in terms of the charge-neutral nature of IEs leading to a self-stabilization effect[20,21] and is consistent with claims of a temperature-independent Mott transition at the lowest temperatures[13]. It is worth noting that the Mott density $n_M$ at which the binding energy vanishes is larger than $n_{IE}^*$, but of the same order[21]. Theoretically, the excitonic Mott density is expected to be given by $n_M a_B^2 \sim 0.1$ in our experiment[13,42,43] (see Supplementary Information S9). Using $a_B \approx 1.9$ nm[11], we estimate $n_M$ ~2.8×10$^{12}$ cm$^{-2}$, providing a consistent upper bound for our measured $n_{eh}^*$ below 20 K. At higher temperatures, we find that $n_{eh}^*(T)$ increases with increasing $T$ with a similar slope to the degeneracy limit defined by $\lambda_T = D$. Here, as $T$ increases, $\lambda_T$ becomes smaller and a higher IE density is required for IE wavefunction overlap to become significant enough to reach the ionization threshold[20–22,24].

## Discussion

We note that in similar systems with an h-BN spacer layer, an opposite trend for the IE phase diagram has been observed[44–47]. However, due to the increased electron-hole separation, the IEs in those systems have a binding energy that is an order of magnitude smaller than in our study. For weakly bound IEs, temperature might be relevant in promoting IE dissociation, resulting in a reduction of the Mott density. In a strongly bound IE system, as studied here, the temperature is always below 10% of the binding energy and ionization physics is dominated by the quantum ionization process described above. Our arguments are supported by a recent exciton drag experiment, which observed a similar $n_{eh}^*(T)$ trend as in our work at the lowest temperatures, before transitioning to a thermal ionization dominated regime[45].

Using a gate-tunable trap architecture, we have demonstrated a novel feature of critical IE dissociation dynamics: an exponential linewidth broadening. This reveals a crucial piece of the IE phase diagram, and we hope future theoretical studies will help address its microscopic origins. Combining these strong many-body effects with highly gate tunable devices paves the way for novel optoelectronic devices such as exciton transistors[10,48,49] or high population density inverted cavity-less lasers[50]. Furthermore, depth-tunable traps are an important step towards funneling a controlled density of cold dipolar excitons away from the excitation spot, thereby eliminating any laser-induced heating or coherence effects that could prevent unambiguous identification of a condensate.

**Acknowledgments:** We thank Ilya Esterlis and Eugene Demler for helpful discussions. P.K. acknowledges the support from the ONR MURI program (N00014-21-1-2377). A.Y.J. is supported by Samsung Electronics. A.M.M.V. is supported by AFOSR (FA2386-21-1-4086). K.W. and T.T. acknowledge support from the JSPS KAKENHI (Grant Numbers 21H05233 and 23H02052) and World Premier International Research Center Initiative (WPI), MEXT, Japan.
**Authors' contributions:** A.Y.J., L.A.J, and P.K. conceived the study. A.Y.J, A.M.M.V., L.A.J. K.P., D.D., Y.Z., G.S., K.D.G. and A.S performed the experiments. A.Y.J., L.A.J., and K.P. fabricated the device. B.K. and J.H. performed the TMD crystal growth. T.T. and K.W. performed the h-BN crystal growth. M.D.L, H.P., and P.K. supervised the experiments. A.Y.J., A.M.M.V., and P.K. analyzed the data and wrote the manuscript with extensive input from all authors.
**Competing financial interests:** The authors declare no competing financial interests.

**References**

1. Chernikov, A. *et al.* Exciton binding energy and nonhydrogenic Rydberg series in monolayer $WS_2$. *Phys. Rev. Lett.* **113**, 076802 (2014).

2. Mak, K. F., He, K., Shan, J. & Heinz, T. F. Control of valley polarization in monolayer $MoS_2$ by optical helicity. *Nat. Nanotechnol.* **7**, 494–498 (2012).

3. Onga, M., Zhang, Y., Ideue, T. & Iwasa, Y. Exciton Hall effect in monolayer $MoS_2$. *Nat. Mater.* **16**, 1193–1197 (2017).

4. Jauregui, L. A. *et al.* Electrical control of interlayer exciton dynamics in atomically thin heterostructures. *Science* **366**, 870–875 (2019).

5. Joe, A. Y. *et al.* Electrically controlled emission from singlet and triplet exciton species in atomically thin light-emitting diodes. *Phys. Rev. B* **103**, L161411 (2021).

6. Rivera, P. *et al.* Observation of long-lived interlayer excitons in monolayer MoSe2–WSe2 heterostructures. *Nat. Commun.* **6**, 6242 (2015).

7. Unuchek, D. *et al.* Valley-polarized exciton currents in a van der Waals heterostructure. *Nat. Nanotechnol.* **14**, 1104–1109 (2019).

8. Nagler, P. *et al.* Giant magnetic splitting inducing near-unity valley polarization in van der Waals heterostructures. *Nat. Commun.* **8**, 1551 (2017).

9. Wang, J. *et al.* Optical generation of high carrier densities in 2D semiconductor heterobilayers. *Sci. Adv.* **5**, eaax0145 (2019).

10. Unuchek, D. *et al.* Room-temperature electrical control of exciton flux in a van der Waals heterostructure. *Nature* **560**, 340–344 (2018).

11. Van Der Donck, M. & Peeters, F. M. Interlayer excitons in transition metal dichalcogenide heterostructures. *Phys. Rev. B* **98**, 115104 (2018).

12. Gillen, R. & Maultzsch, J. Interlayer excitons in MoSe2/WSe2 heterostructures from first principles. *Phys. Rev. B* **97**, 165306 (2018).

13. Fogler, M. M., Butov, L. V. & Novoselov, K. S. High-temperature superfluidity with indirect excitons in van der Waals heterostructures. *Nat. Commun.* **5**, 4555 (2014).

14. High, A. A. *et al.* Spontaneous coherence in a cold exciton gas. *Nature* **483**, 584–588 (2012).

15. High, A. A. *et al.* Condensation of excitons in a trap. *Nano Lett.* **12**, 2605–2609 (2012).

16. Alloing, M. *et al.* Evidence for a Bose-Einstein condensate of excitons. *EPL Europhys. Lett.* **107**, 10012 (2014).

17. Rossbach, G. *et al.* High-temperature Mott transition in wide-band-gap semiconductor quantum wells. *Phys. Rev. B* **90**, 201308 (2014).

18. Stern, M., Garmider, V., Umansky, V. & Bar-Joseph, I. Mott Transition of Excitons in Coupled Quantum Wells. *Phys. Rev. Lett.* **100**, 256402 (2008).

19. Kiršanskė, G. *et al.* Observation of the exciton Mott transition in the photoluminescence of coupled quantum wells. *Phys. Rev. B* **94**, 155438 (2016).

20. Snoke, D. Predicting the ionization threshold for carriers in excited semiconductors. *Solid State Commun.* **146**, 73–77 (2008).

21. Asano, K. & Yoshioka, T. Exciton–Mott Physics in Two-Dimensional Electron–Hole Systems: Phase Diagram and Single-Particle Spectra. *J. Phys. Soc. Jpn.* **83**, 084702 (2014).

22. Reinholz, H. Mott effect for an electron–hole plasma in a two-dimensional structure. *Solid State Commun.* **123**, 489–494 (2002).

23. Nikolaev, V. V. & Portnoi, M. E. Theory of excitonic Mott transition in double quantum wells. *Phys. Status Solidi C* **1**, 1357–1362 (2004).

24. Manzke, G., Semkat, D. & Stolz, H. Mott transition of excitons in GaAs-GaAlAs quantum wells. *New J. Phys.* **14**, 095002 (2012).

25. Zimmermann, R. 5.2 The Mott transition of excitons. in *Many-Particle Theory of Highly Excited Semiconductors* 119–143 (B. G. Teubner Verlagsgesellschaft, Leipzig, 1988).

26. Wang, Z. *et al.* Evidence of high-temperature exciton condensation in two-dimensional atomic double layers. *Nature* **574**, 76–80 (2019).

27. Sigl, L. *et al.* Signatures of a degenerate many-body state of interlayer excitons in a van der Waals heterostack. *Phys. Rev. Res.* **2**, 042044 (2020).

28. Sushko, A. *et al.* Asymmetric photoelectric effect: Auger-assisted hot hole photocurrents in transition metal dichalcogenides. *Nanophotonics* **10**, 105–113 (2020).

29. Binder, J. *et al.* Upconverted electroluminescence via Auger scattering of interlayer excitons in van der Waals heterostructures. *Nat. Commun.* **10**, 2335 (2019).

30. Choi, J. *et al.* Moiré potential impedes interlayer exciton diffusion in van der Waals heterostructures. *Sci. Adv.* **6**, eaba8866 (2020).

31. Shanks, D. N. *et al.* Nanoscale Trapping of Interlayer Excitons in a 2D Semiconductor Heterostructure. *Nano Lett.* **21**, 5641–5647 (2021).

32. Seyler, K. L. *et al.* Signatures of moiré-trapped valley excitons in MoSe2/WSe2 heterobilayers. *Nature* **567**, 66–70 (2019).

33. Andersen, T. I. *et al.* Excitons in a reconstructed moiré potential in twisted WSe2/WSe2 homobilayers. *Nat. Mater.* **20**, 480–487 (2021).

34. Li, W., Lu, X., Dubey, S., Devenica, L. & Srivastava, A. Dipolar interactions between localized interlayer excitons in van der Waals heterostructures. *Nat. Mater.* **19**, 624–629 (2020).

35. Kremser, M. *et al.* Discrete interactions between a few interlayer excitons trapped at a MoSe2–WSe2 heterointerface. *Npj 2D Mater. Appl.* **4**, 8 (2020).

36. Ciarrocchi, A. *et al.* Polarization switching and electrical control of interlayer excitons in two-dimensional van der Waals heterostructures. *Nat. Photonics* **13**, 131–136 (2019).

37. Wang, J. *et al.* Diffusivity Reveals Three Distinct Phases of Interlayer Excitons in $MoSe_2/WSe_2$ Heter. *Phys. Rev. Lett.* **126**, 106804 (2021).

38. Negoita, V., Snoke, D. & Eberl, K. Huge density-dependent blueshift of indirect excitons in biased coupled quantum wells. *Phys. Rev. B - Condens. Matter Mater. Phys.* **61**, 2779–2783 (2000).

39. Butov, L. V., Shashkin, A. A., Dolgopolov, V. T., Campman, K. L. & Gossard, A. C. Magneto-optics of the spatially separated electron and hole layers in ${\mathrm{G}\mathrm{a}\mathrm{A}\mathrm{s}/\mathrm{A}\mathrm{l}}_{x}{\mathrm{Ga}}_{1\ensuremath{-}x}\mathrm{As}$ coupled quantum wells. *Phys. Rev. B* **60**, 8753–8758 (1999).

40. Laikhtman, B. & Rapaport, R. Exciton correlations in coupled quantum wells and their luminescence blue shift. *Phys. Rev. B - Condens. Matter Mater. Phys.* **80**, (2009).

41. Kappei, L., Szczytko, J., Morier-Genoud, F. & Deveaud, B. Direct observation of the Mott transition in an optically excited semiconductor quantum well. *Phys. Rev. Lett.* **94**, 1–4 (2005).

42. Wu, F.-C., Xue, F. & MacDonald, A. H. Theory of two-dimensional spatially indirect equilibrium exciton condensates. *Phys. Rev. B* **92**, 165121 (2015).

43. De Palo, S., Rapisarda, F. & Senatore, G. Excitonic Condensation in a Symmetric Electron-Hole Bilayer. *Phys. Rev. Lett.* **88**, 206401 (2002).

44. Ma, L. *et al.* Strongly correlated excitonic insulator in atomic double layers. *Nature* **598**, 585–589 (2021).

45. Nguyen, P. X. *et al.* Perfect Coulomb drag in a dipolar excitonic insulator. Preprint at https://doi.org/10.48550/arXiv.2309.14940 (2023).

46. Qi, R. *et al.* Perfect Coulomb drag and exciton transport in an excitonic insulator. Preprint at https://doi.org/10.48550/arXiv.2309.15357 (2023).

47. Qi, R. *et al.* Thermodynamic behavior of correlated electron-hole fluids in van der Waals heterostructures. *Nat. Commun.* **14**, 8264 (2023).

48. High, A. A., Novitskaya, E. E., Butov, L. V., Hanson, M. & Gossard, A. C. Control of Exciton Fluxes in an Excitonic Integrated Circuit. *Science* **321**, 229–231 (2008).

49. Liu, Y. *et al.* Electrically controllable router of interlayer excitons. *Sci. Adv.* **6**, eaba1830 (2020).

50. Paik, E. Y. *et al.* Interlayer exciton laser of extended spatial coherence in atomically thin heterostructures. *Nature* **576**, 80–84 (2019).

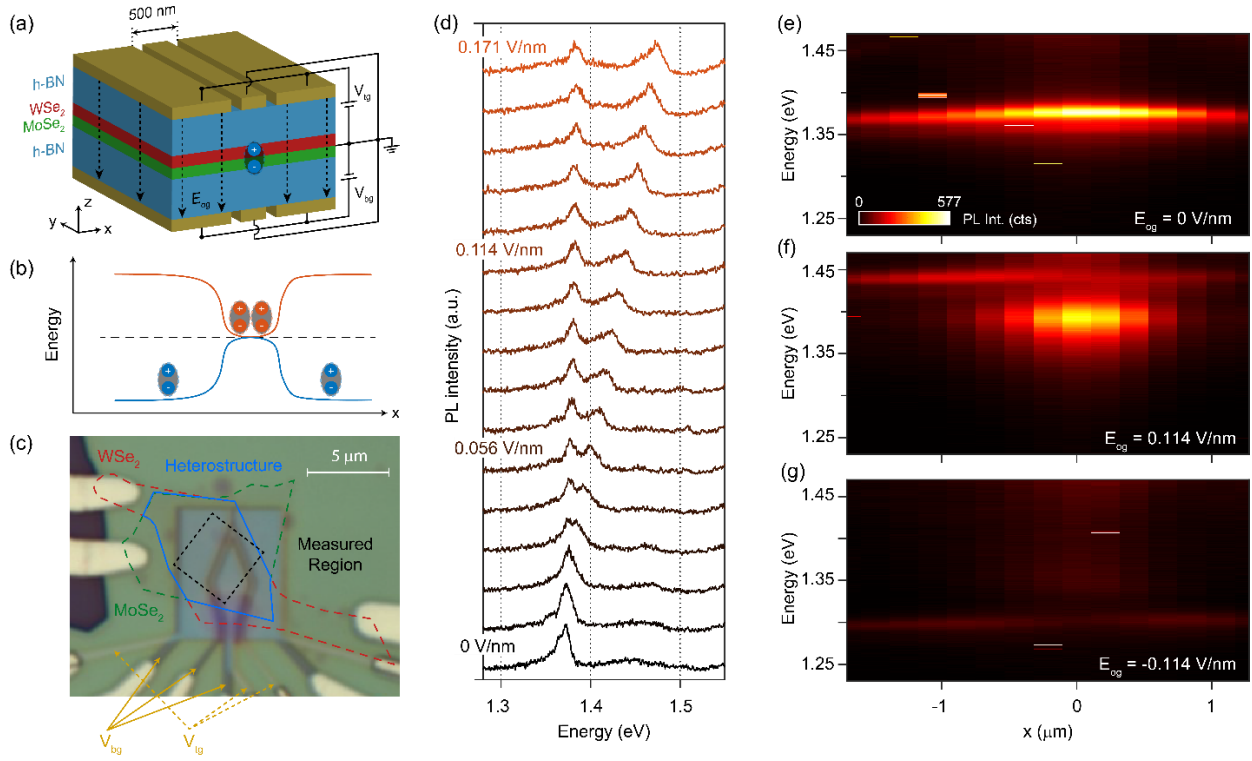

**Fig. 1 Modulating the interlayer exciton energy with electric fields** (a) Device schematic showing the stripe gate and large outer gates on either side. $V_{tg}$ and $V_{bg}$ are applied to create electric fields in the outer gate ($E_{og}$) regions while the stripe gates are grounded. (b) IE energy schematic as a function of *x*-position perpendicular to the stripe gate. The red (black dashed) curves represent the electric field profile felt by the IEs with (without) $E_{og}$ in a trapping configuration. The blue curve represents the anti-trap regime. (c) Optical image of the device. Red and green dashed lines outline the WSe$_2$ and MoSe$_2$ regions, respectively. Blue line outlines the heterostructure region. Gold arrows identify the relevant top gate and bottom gate structures. The black dashed square indicates the approximate region for the PL diffusion measurements. (d) IE spectra as a function of $E_{og}$. The linearly shifting peak at higher energy is from the outer gate region while the lower energy peak is from within the stripe gate region. (e)-(g) Scanning PL spectroscopy, where the laser excites at *x* = 0 μm and the spectra is collected across the stripe gate at $E_{og}$ = 0 and ±0.114 V/nm, showing spatially modulated IE energy.

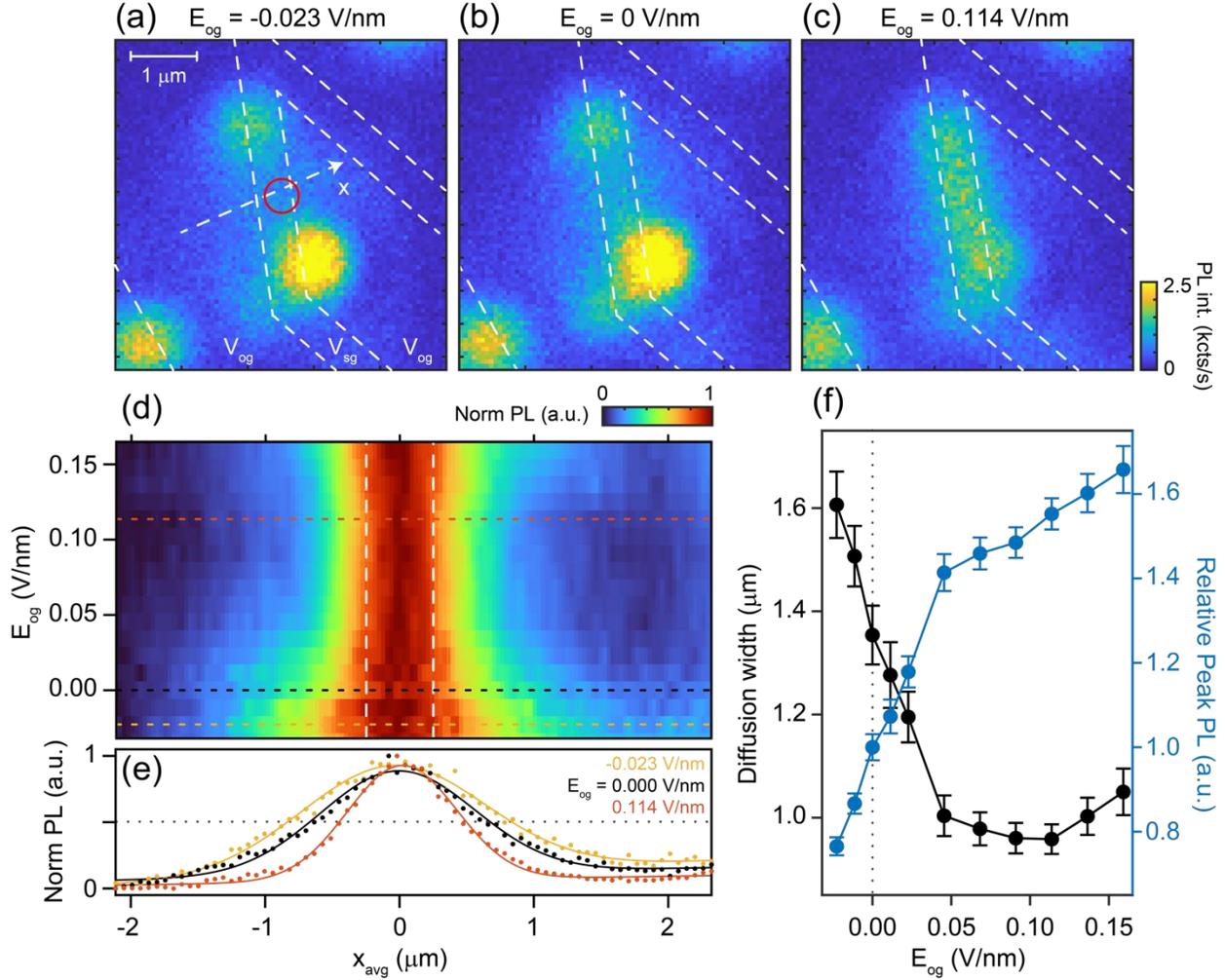

**Fig. 2 Controlling the diffusion behavior** (a)-(c) Spatial mapping of the diffused IE PL emission at $E_{og}$ = −0.023, 0, and 0.114 V/nm. The outer gate and the stripe gate areas are labeled $V_{og}$ and $V_{sg}$, respectively. White dashed arrow indicates positive $x$ direction. Red circle is the excitation position. Bright spots that naturally occur on the sample are reduced with the trapping potential. (d) Normalized PL along the $x$ direction, averaged over the uniform $y$-cut trap region, as a function of $E_{og}$. The dotted lines correspond to the $E_{og}$ line cuts in (e). The white dashed lines indicate the expected trap width of ∼ 500nm. (e) Line cuts (scatter) and Gaussian fittings (lines) at $E_{og}$ = −0.023, 0, and 0.114 V/nm showing narrowing (broadening) of the diffusion width when forming a trap (anti-trap). (f) Extracted diffusion width and relative peak intensity ($PL_{peak}(E_{og})/PL_{peak}(E_{og} = 0)$) as a function of $E_{og}$. We observe the trap saturates near $E_{og} \approx 0.04$ V/nm.

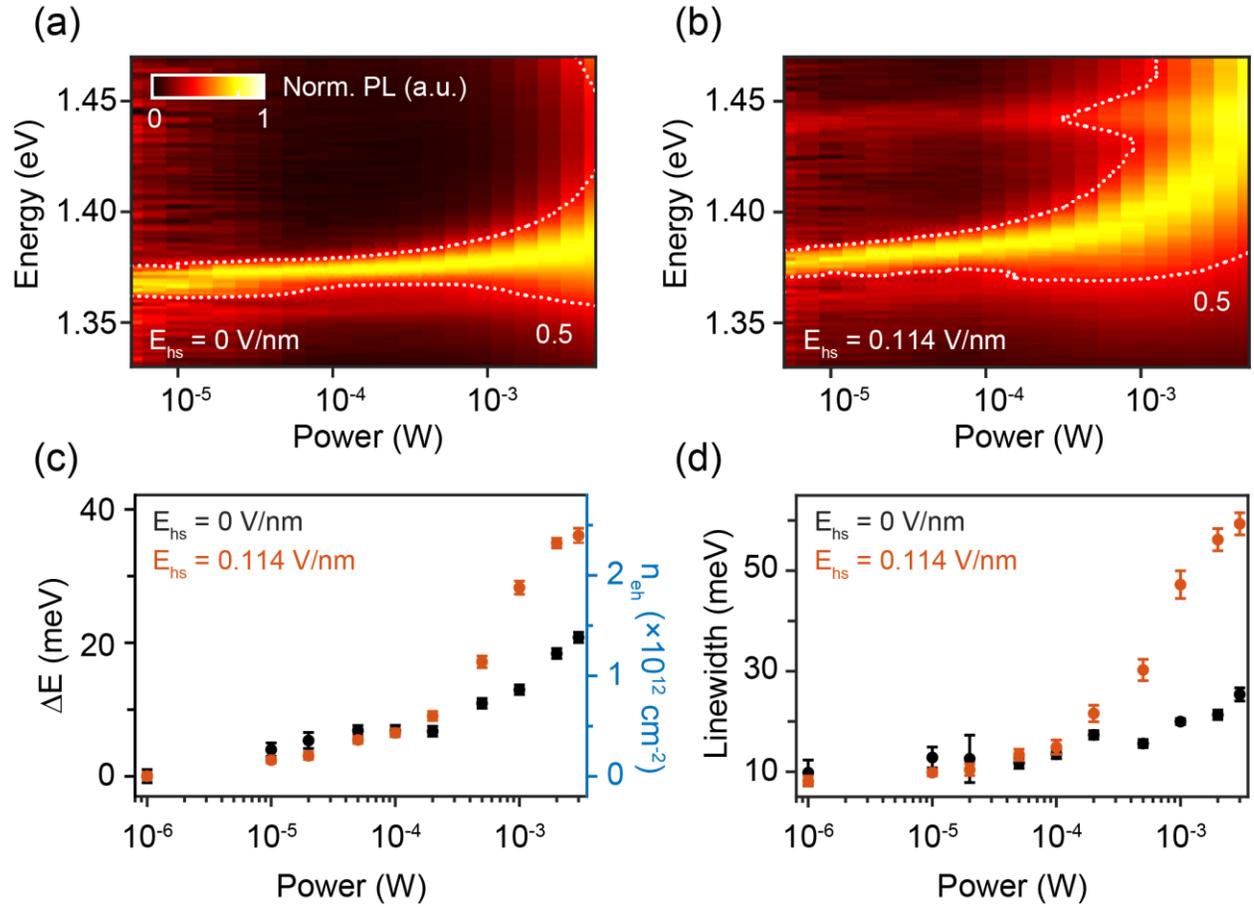

**Fig. 3 Tuning the interlayer exciton density** (a) Normalized power dependent PL emission from the stripe gate region at $E_{og}$ = 0 V/nm, when the excitation laser is slightly outside the region. The dotted white line shows a contour of 0.5 for the intensity of the trapped IEs. (b) Same as (a) for $E_{og}$ = 0.114 V/nm. (c) The blueshift energy, $\Delta E$, for $E_{og}$ = 0 and 0.114 V/nm as a function of power. Right axis (blue) shows the corresponding calculated electron-hole pair density, $n_{eh}$. (d) Fitted trapped PL linewidth for $E_{og}$ = 0 and 0.114 V/nm as a function of power. The error bars in (c) and (d) are from the fitting of the spectral peaks.

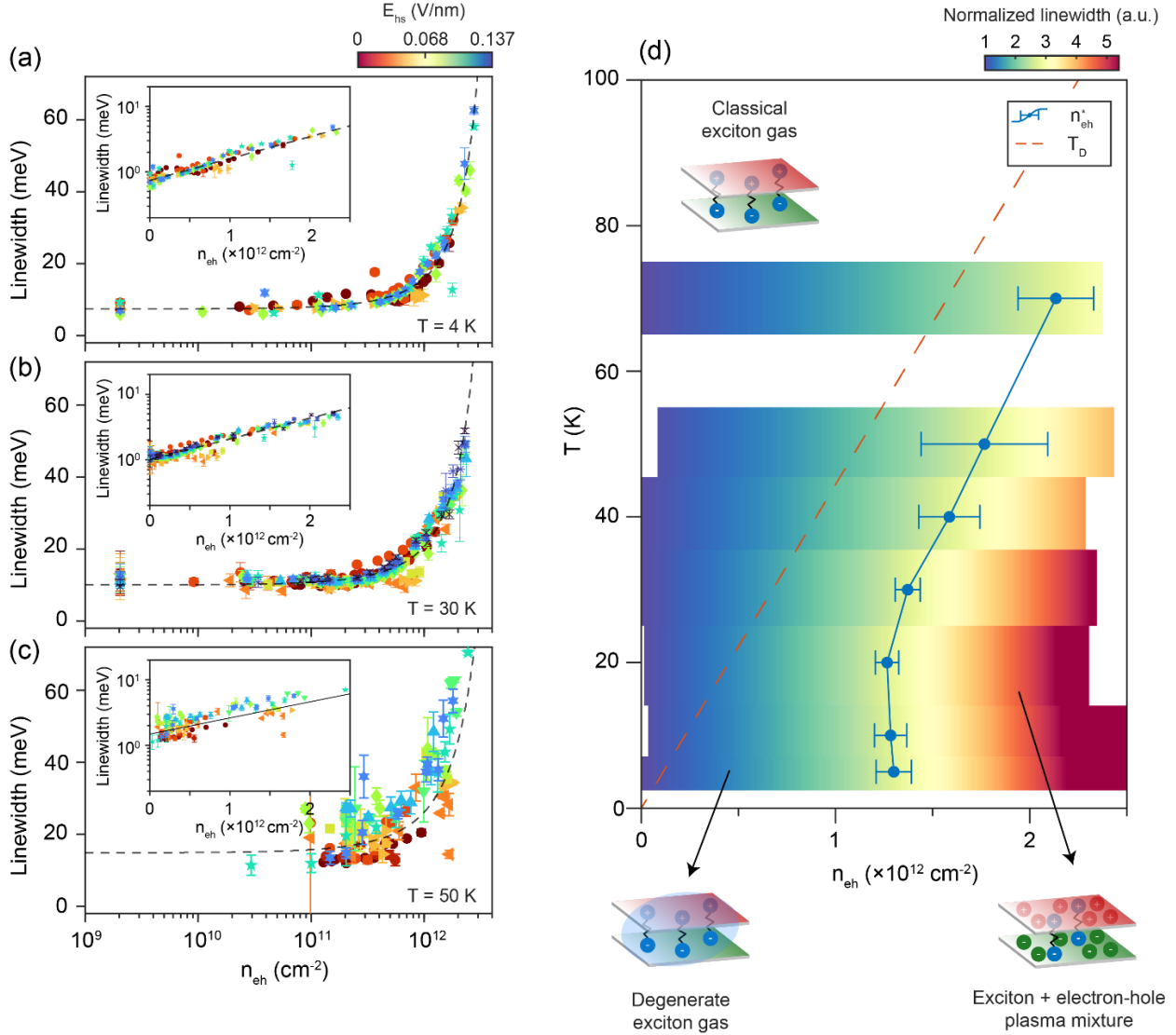

**Fig. 4 Interlayer exciton phase diagram** (a) Scatter plot of linewidth vs. electron-hole density for various trap depths at T = 4 K. The error bars indicate the uncertainty of the fittings. The black dashed line shows the exponential fit giving a universal critical density $n^*_{eh}(4K) = 1.30(\pm 0.09) \times 10^{12}$ cm$^{-2}$. Inset: Same data but where linewidth is on a log scale and density on linear scaling, showing the exponential fit is shown as a linear curve. (b-c) Same as (a) but at 30 K with $n^*_{eh}(30K) = 1.37(\pm 0.06) \times 10^{12}$ cm$^{-2}$ and 50 K with $n^*_{eh}(30K) = 1.77(\pm 0.32) \times 10^{12}$ cm$^{-2}$, respectively. (d) Plot of the fitted normalized linewidth over the collected data range as a function of pair density and measured temperature. The blue data shows the extracted $n^*_{eh}(T)$ at different temperatures with the error bars showing the uncertainty of the fits. The red dashed line is the degeneracy temperature, $T_D$ as a function of exciton density per flavor.

# Methods

**Device fabrication**

Figure 1(c) shows our h-BN encapsulated $MoSe_2/WSe_2$ heterostructure device D1 with electrical contacts to each layer. The $MoSe_2$, $WSe_2$, and h-BN layers are prepared via mechanical exfoliation on 285 nm $SiO_2$ substrates. The $MoSe_2$ and $WSe_2$ bulk crystals are grown by the flux method. The h-BN thicknesses are 36 nm and 59 nm for the top and bottom h-BN, respectively. The stack is assembled by picking up each layer starting with the top h-BN using a standard dry transfer method. The final stack is placed on ultra-flat Cr/PdAu alloy (1 nm / 9 nm) bottom gates written on a 285 nm $SiO_2$ substrate. The stripe gate width (~ 250 nm) and outer gate separations (~ 500 nm) are measured via scanning electron microscopy images (Extended Data Fig. 1). The same gates are evaporated on top of the completed heterostructure. Edge contacts were fabricated by reactive ion etching parts of the monolayer region with $O_2/CHF_3/Ar$ gas mixture and then evaporating Cr / Au leads (5 nm / 120 nm). It should be noted these contacts are not sufficient for electronic transport but can be used for electrostatic doping measurements.

**Measurement details**

Optical measurements were performed using a 4f confocal microscope system with a 0.75 NA 100x objective in a Montana Instruments, closed-loop optical 4 K cryostat. Measurements were performed at $T = 4$ K, except for temperature dependent measurements. PL measurements were performed with a Thorlabs continuous wave diode laser with excitation wavelength of 660 nm and collected into a spectrometer with a PIXIS:256BR or BLAZE:400HR camera (both Princeton Instruments) unless otherwise noted. Scanning images were taken by fiber coupling into an avalanche photodiode (Excelitas SPCM-900-14-FC). Time resolved PL measurements are performed with a supercontinuum laser from NKT with an 80 MHz repetition rate and through a tunable bandpass filter (SuperK VARIA). The laser is synchronized to a single photon counting module (Excelitas Technologies) and a picosecond event timer (Picoharp 300, PicoQuant). Scanning confocal photoluminescence spectra are taken by using two separate optical paths to excite and collect from separate positions. The optical path is split using a 50:50 pellicle beamsplitter, each with a set of controllable two-axis galvanometer mirrors. The excitation channel has a co-aligned collection line that can be used to choose the excitation position based on PL map. With the excitation fixed, the collection channel is scanned to measure PL emission counts or spectra from the various parts of the sample. All PL measurements presented in the manuscript are measured this way unless stated otherwise. The PL map in Fig. S2(b) is taken where the excitation and collection are co-aligned. All electrostatic gates are controlled via Keithley 2400 sourcemeters.

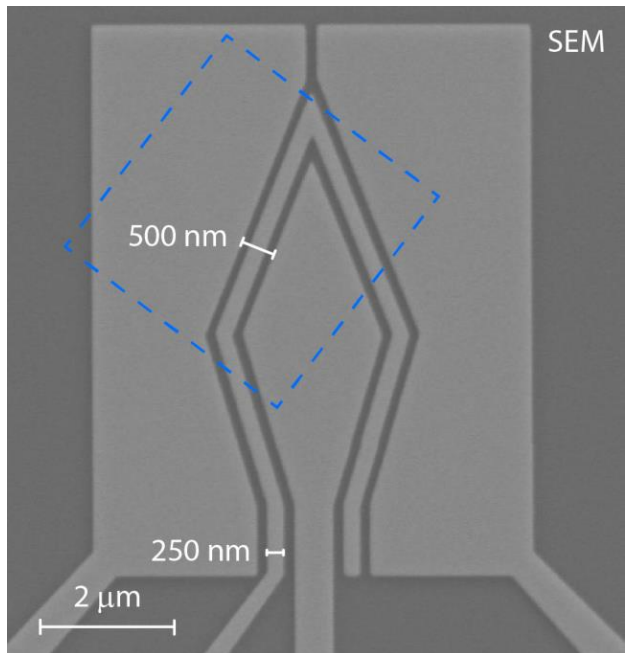

**Extended Data Fig. 1:** SEM image of the Cr/PdAu bottom gates with the stripe gate width and outer gate widths identified. The blue dashed square indicates the approximate location for the PL diffusion measurements.